\documentclass[fleqn,usenatbib]{mnras}
\usepackage{newtxtext,newtxmath}
\usepackage[T1]{fontenc}
\usepackage{ae,aecompl}
\usepackage{graphicx}	
\usepackage{amsmath}	
\usepackage{amssymb}	

\title[Radio timing observations of XTE J1810--197]{Spin frequency evolution and pulse profile variations of the recently re-activated radio magnetar XTE J1810--197}

\author[L. Levin et al.]{
L. Levin,$^{1}$\thanks{E-mail: Lina.Preston@manchester.ac.uk}
A. G. Lyne,$^{1}$
G. Desvignes,$^{2}$
R. P. Eatough,$^{2}$
R. Karuppusamy,$^{2}$
\newauthor 
M. Kramer,$^{2,1}$
M. Mickaliger,$^{1}$
B. W. Stappers,$^{1}$
and P. Weltevrede$^{1}$
\\
$^{1}$Jodrell Bank Centre for Astrophysics, School of Physics and Astronomy, University of Manchester, Manchester M13 9PL, UK\\
$^{2}$Max-Planck-Institut f{\"u}r Radioastronomie, Auf dem H{\"u}gel 69, D-53121 Bonn, Germany
}

\date{Accepted XXX. Received YYY; in original form ZZZ}

\pubyear{2019}

\begin{document}
\label{firstpage}
\pagerange{\pageref{firstpage}--\pageref{lastpage}}
\maketitle

\begin{abstract}
After spending almost a decade in a radio-quiet state, the Anomalous X-ray Pulsar XTE\,J1810--197 turned back on in early December 2018. We have observed this radio magnetar at 1.5 GHz with $\sim$daily cadence since the first detection of radio re-activation on 8 December 2018. In this paper, we report on the current timing properties of XTE\,J1810--197 and find that the magnitude of the spin frequency derivative has increased by a factor of 2.6 over our 48-day data set. We compare our results with the spin-down evolution reported during its previous active phase in the radio band. We also present total intensity pulse profiles at five different observing frequencies between 1.5 and 8.4 GHz, collected with the Lovell and the Effelsberg telescopes. The profile evolution in our data set is less erratic than what was reported during the previous active phase, and can be seen varying smoothly between observations. Profiles observed immediately after the outburst show the presence of at least five cycles of a very stable $\sim$50-ms periodicity in the main pulse component that lasts for at least tens of days. This remarkable structure is seen across the full range of observing frequencies.
\end{abstract}



\section{Introduction}

Magnetars are slow-spinning neutron stars with extremely high surface magnetic field strengths. Their emission is thought to be powered by the decay of their magnetic fields \citep{dun92}, and they are known to occasionally undergo very bright X-ray outbursts. XTE\,J1810--197 is one of 23 currently known magnetars\footnote{http://www.physics.mcgill.ca/$\sim$pulsar/magnetar/main.html}, and was the first of only four known to emit radio pulsations \citep{cam06Nat}.   
It has a spin period of $\sim$5.54\,s and an inferred surface magnetic field strength of $\sim$2$\times$10$^{14}$\,G. 
The radio emission from XTE\,J1810--197 was detected a year after an X-ray outburst in 2003 
\citep{hal05} 
and was observed to fade as the X-ray emission faded for a few years after the outburst. 
During this time, the magnetar displayed pulsed radio emission that was highly variable, both with respect to integrated pulse 
profiles \citep{camApJ663}, to the single pulse parameters \citep{ser09} and to spin down parameters \citep{camApJ663}. 
The emission exhibited nearly 100$\%$ linear polarisation, which could be seen at a range of observing frequencies in both the integrated pulse profiles and in single pulses \citep{kra07,camApJ659}. 
Also the spectral index of the radio emission fluctuated with time \citep{camApJ669, laz08}. 
The X-ray flux returned to pre-outburst values in 2007-2008 \citep{ber11} and the radio pulsations dropped below the radio detection threshold in late 2008 \citep{cam16}.  

On 8 Dec 2018, we detected a bright pulsed radio signal at 1.52 GHz from the magnetar XTE\,J1810--197 with the Lovell telescope at Jodrell Bank Observatory (JBO), as part of a regular monitoring program of this source \citep{lyn18atel}. 
This detection marked the end of a decade of radio-quietness, and XTE\,J1810--197 has been detected with high flux density in all our subsequent observations. 
It is unclear exactly when the radio emission was re-activated. The last non-detection in the radio band collected at JBO occurred on 26 Oct 2018. In the X-ray band, the MAXI telescope has detected emission from XTE\,J1810--197 at 2-10 keV since 26 Nov 2018, with the last non-detection collected on 20 Nov 2018 \citep{got19}. Subsequently, also the NuSTAR telescope detected pulsed X-ray emission from the magnetar, with flux up to at least 30 keV, in an observation on 13 Dec 2018 \citep{got18atel}. 
The NuSTAR observation shows an absorbed 2-10 keV flux of $(2.12\pm0.07)\times10^{-10}$\,erg/s/cm$^2$ \citep{got19}, which is twice as bright as the maximum possible X-ray flux from 2003, when the magnetar was last observed in outburst, calculated by fitting an exponential decay model and extrapolating back to the last non-detection \citep{got07}. 

Following the 1.52-GHz JBO detection, XTE\,J1810--197 has been detected also at a number of other radio telescopes at centre frequencies ranging from 650\,MHz up to 32\,GHz \citep[e.g.][]{des18atel, low18atel, jos18atel, maj19atel}. 
\cite{dai19} report on three observations using the Ultra Wideband Low receiver at the Parkes Radio Telescope, that show large variation in the polarised emission, profile evolution over the band, and a flat radio spectrum. 


In this paper, we will describe the pulse profile variations and timing parameters derived from data collected with the Lovell telescope and the Effelsberg 100-m telescope since the first detection of the re-activation of XTE\,J1810--197.

\section{Observations}
In anticipation of a radio re-activation, we have monitored XTE\,J1810--197 since the beginning of 2009, 
when possible with approximately monthly 30-minute integrations, using the 76-m Lovell telescope at JBO. 
After the re-activation was detected, observations with the Lovell telescope were carried out on a $\sim$daily cadence, with 42 observations spanning 47 days. The length of each observation varied from 20\,minutes to several hours, but most (34 of 42) of the observations were between 30 and 60 minutes long. The data were collected over a 384-MHz wide band centered at 1520 MHz and divided into 768 frequency channels. Each observation was then folded and dedispersed online with an initial timing ephemeris, which was precise enough that no significant smearing was observed across each 20-second long sub-integration. Archive files with 1024 phase bins across the pulse period were written out with all four Stokes parameters and 32 frequency channels. The data were manually inspected and cleaned of radio-frequency interference. 
Flux density values (plotted in Fig. \ref{fig:residuals}) were measured by calibrating the raw on-pulse powers with the system equivalent flux density (SEFD) and the sky temperature in the direction of the magnetar. The SEFD for JBO was measured as a function of telescope elevation and applied to each observation. The sky temperature was estimated as 11.2\,K using the \cite{has81} sky map at 408\,MHz and extrapolated to our centre frequency of 1520\,MHz, assuming a spectral index of -2.55 \citep{law87}. 

Observations with the Effelsberg 100-m telescope of the Max Planck Institute for Radio Astronomy in Germany were collected using four different receivers with centre frequencies (and bandwidth) of 2.25 GHz (500 MHz), 4.85 GHz (500 MHz), 6.0 GHz (4000 MHz), and 8.35 GHz (500 MHz). At all frequencies, all four Stokes parameters were recorded across multiple (from 128 to 4096) frequency channels. We have collected 12 observations on 6 separate days, each typically 10--20 minutes long.

Interstellar scattering is not expected to affect the pulsar signal significantly at these observing frequencies. This is confirmed by the NE2001 Electron Density Model \cite{cor02}, which predicts a pulse broadening of 40\,$\mu$s at our lowest centre frequency of 1.52\,GHz, using the position and dispersion measure (DM) of XTE\,J1810--197 (see table \ref{tab:freq}). 

\section{Profile evolution}

\begin{figure}
	\includegraphics[width=\columnwidth]{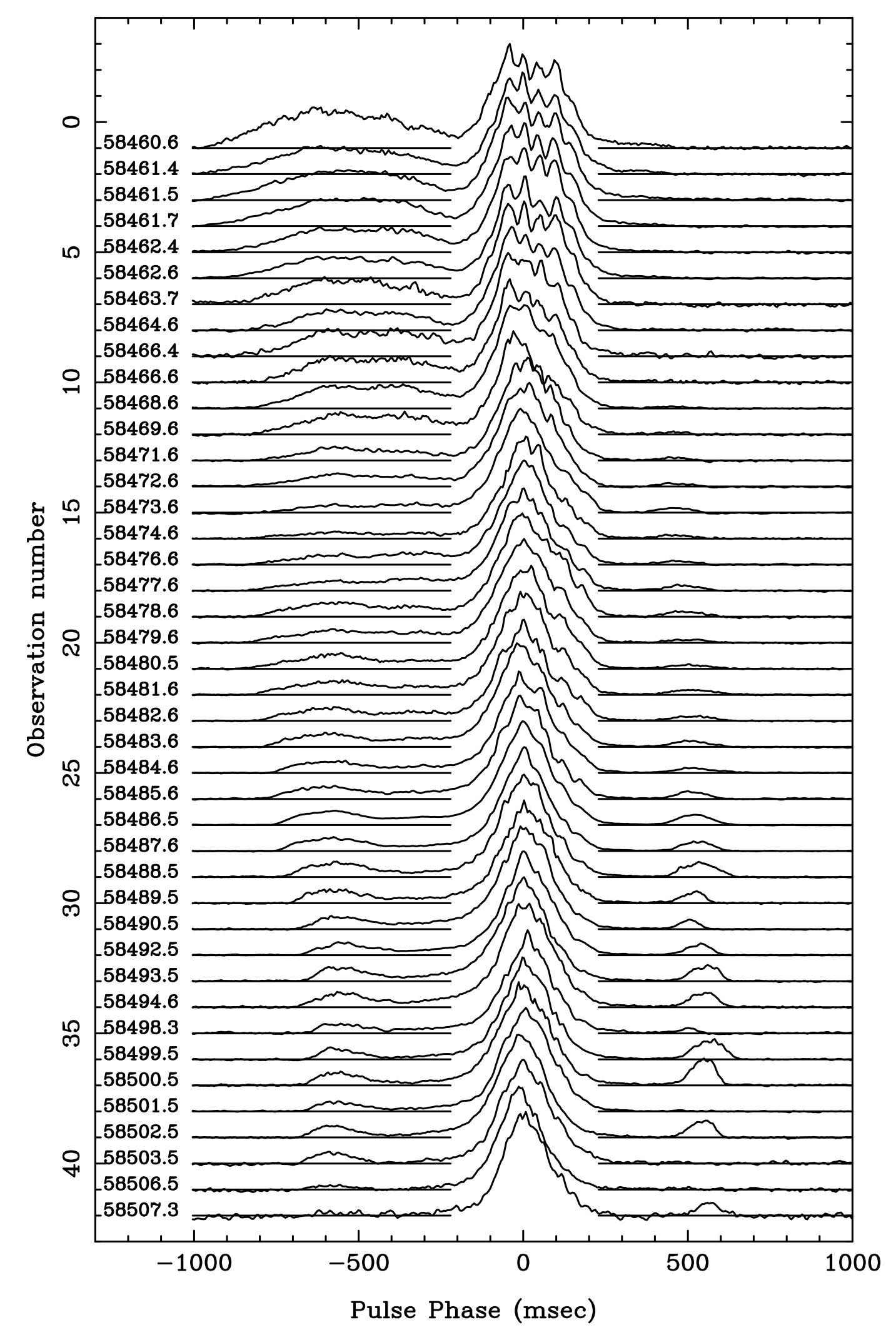}
    \caption{Normalised total intensity profiles for all observations collected with the Lovell telescope at 1.52\,GHz, using 1024 pulse phase bins and displaying 35\% of the rotation. The profiles have been aligned using the timing parameters as reported in Table \ref{tab:freq}. Note the presence of a stable periodic structure on the top of the main (brightest) component during the first 10 observations.}
    \label{fig:profiles}
\end{figure}

\begin{figure}
	\includegraphics[width=\columnwidth]{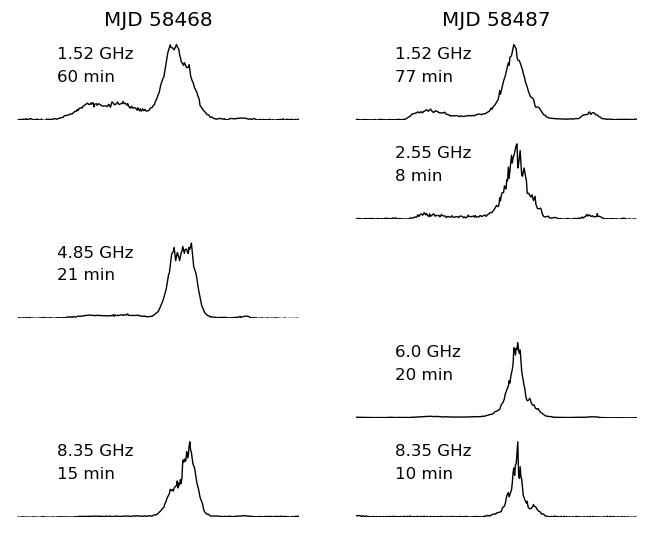}
    \caption{Frequency evolution of the total intensity pulse profiles on MJD 58468 and MJD 58487. The 1.52-GHz data were collected with the Lovell telescope and the higher frequencies with the Effelsberg telescope. At Effelsberg, the observations on each day were collected consecutively, and overlapping in time with at least part of the JBO observations. All observations were created with 1024 pulse phase bins and 35\% of the rotation is displayed. The profiles have been aligned using the timing parameters as reported in Table \ref{tab:freq}.}
    \label{fig:eff-jbo}
\end{figure}

The pulse profile of XTE\,J1810--197 has changed significantly since detection. The total intensity profiles at 1.52 GHz from JBO are shown in Fig \ref{fig:profiles}, where each profile has been normalised by its maximum power. 
Compared to what was observed the last time XTE\,J1810--197 was active in the radio band \citep{camApJ663,kra07}, the profile changes observed during the first two months since the re-activation are less extreme. The temporal evolution of the pulse components can be tracked in our daily observations and are seen to grow and fade with time and drift in relative position. The main peak of the profile has become narrower during the time since it was first redetected, with W$_{50}$=4.4\% on MJD\,58460.6 reducing to W$_{50}$=2.7\% on MJD=58507.3, where W$_{50}$ is the pulse width expressed as percentage of the pulse period at 50\% of the peak intensity. 
At the start of the observations, the pre-cursor component was wider than the main pulse and with peak intensity approximately one third of the main peak. The relative intensity of the pre-cursor to the main peak, and the width of the pre-cursor have both reduced with time, and it is barely visible in the last few observations. 
A narrow post-cursor component appeared around MJD\,58468 and grew in intensity over the next $\sim$20 days. Towards the end of the data span, this post-cursor seems to be intermittent, being bright in some observations and completely absent in others (see e.g. MJDs\,58500.5 and 58501.5 in Fig. \ref{fig:profiles}). 

A comparison of the pulse profiles over frequency on two days is shown in Fig \ref{fig:eff-jbo}. 
Here we can see that the main pulse component becomes narrower at higher observing frequency, and different parts of the main pulse are brighter at different frequencies (see e.g. MJD 58468 at 1.52 GHz compared to 8.35 GHz). Both the pre-cursor and the post-cursor are relatively strong at 1.52 GHz, and get weaker at higher observing frequencies. Neither is visible in the 8.35 GHz data. 
When making these comparisons, it is important to note that the Effelsberg observations in general are shorter than those collected at the JBO, and hence have sampled fewer rotations of the magnetar. The length of each observation is noted on the relevant panel in Fig \ref{fig:eff-jbo}. 
A deeper analysis of the high frequency data, including a study of the polarisation properties of all Effelsberg and JBO data, will be published in Desvignes et al ({\it in prep}).

\subsection{Periodicity within main profile component}
\begin{figure}
    \centering
    \includegraphics[width=\columnwidth]{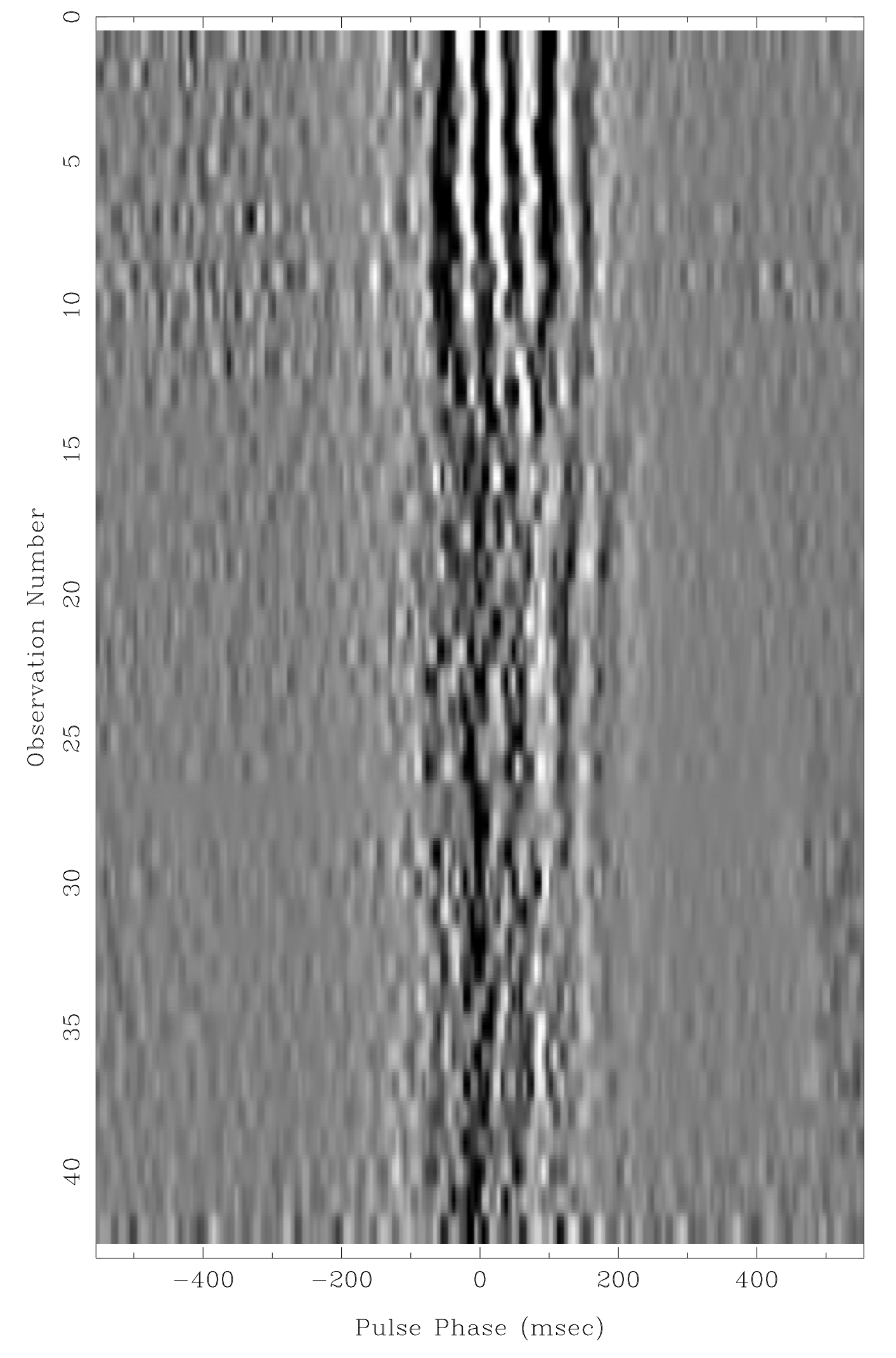}
    \caption{A grey-scale plot of the region around the main pulsed component at 1.52 GHz after the application of a high-pass filter to the data of Fig. \ref{fig:profiles}.  
    In this diagram, peaks in flux density are represented by dark shades, troughs by lighter shades. The stability and strength of the oscillations are clear during the first dozen observations ($\sim$10 days), after which they decrease.}
    \label{fig:wiggles}
\end{figure}

Close inspection of the first few JBO observations (Fig. \ref{fig:profiles}) shows that the main profile component has a flat top with a superposed periodic structure. This pattern has a period of $\sim$50\,ms and is stable in phase and amplitude even between observations. 
This high-frequency structure is revealed more clearly by application of a high-pass filter to the profiles presented in Fig. \ref{fig:profiles} to remove the obfuscating broad structure of the pulses. 
The filter was achieved by the removal of a gaussian band of fluctuation frequencies having a half-power width of 13.8 Hz. 
Fig. \ref{fig:wiggles} shows a grey-scale image of the filtered data, showing that the periodicity is stable in strength and phase for about 10 days, before becoming weaker, less stable in phase and less distinct.

\begin{figure*}
    \includegraphics[width=7cm]{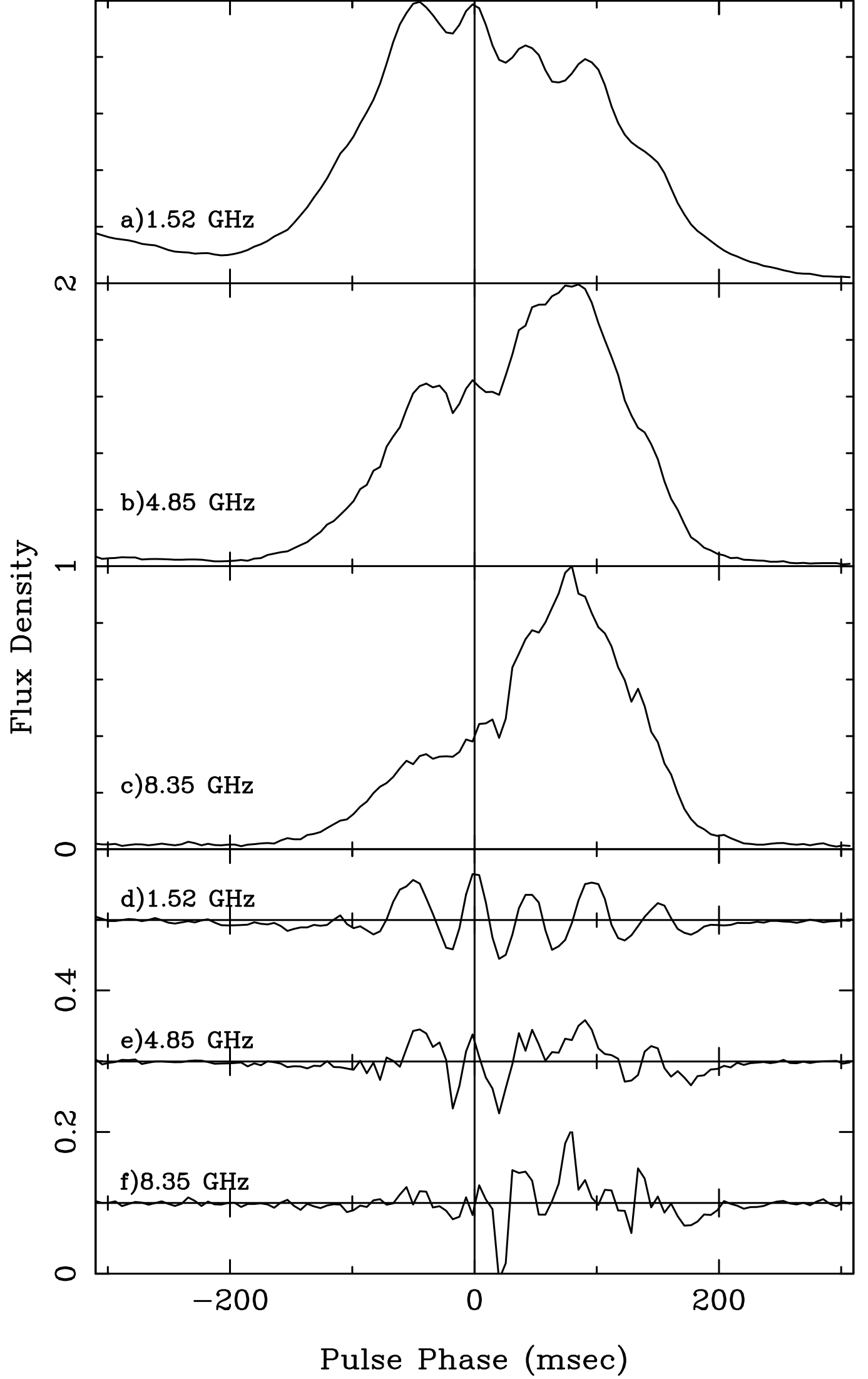}
	\includegraphics[width=7cm]{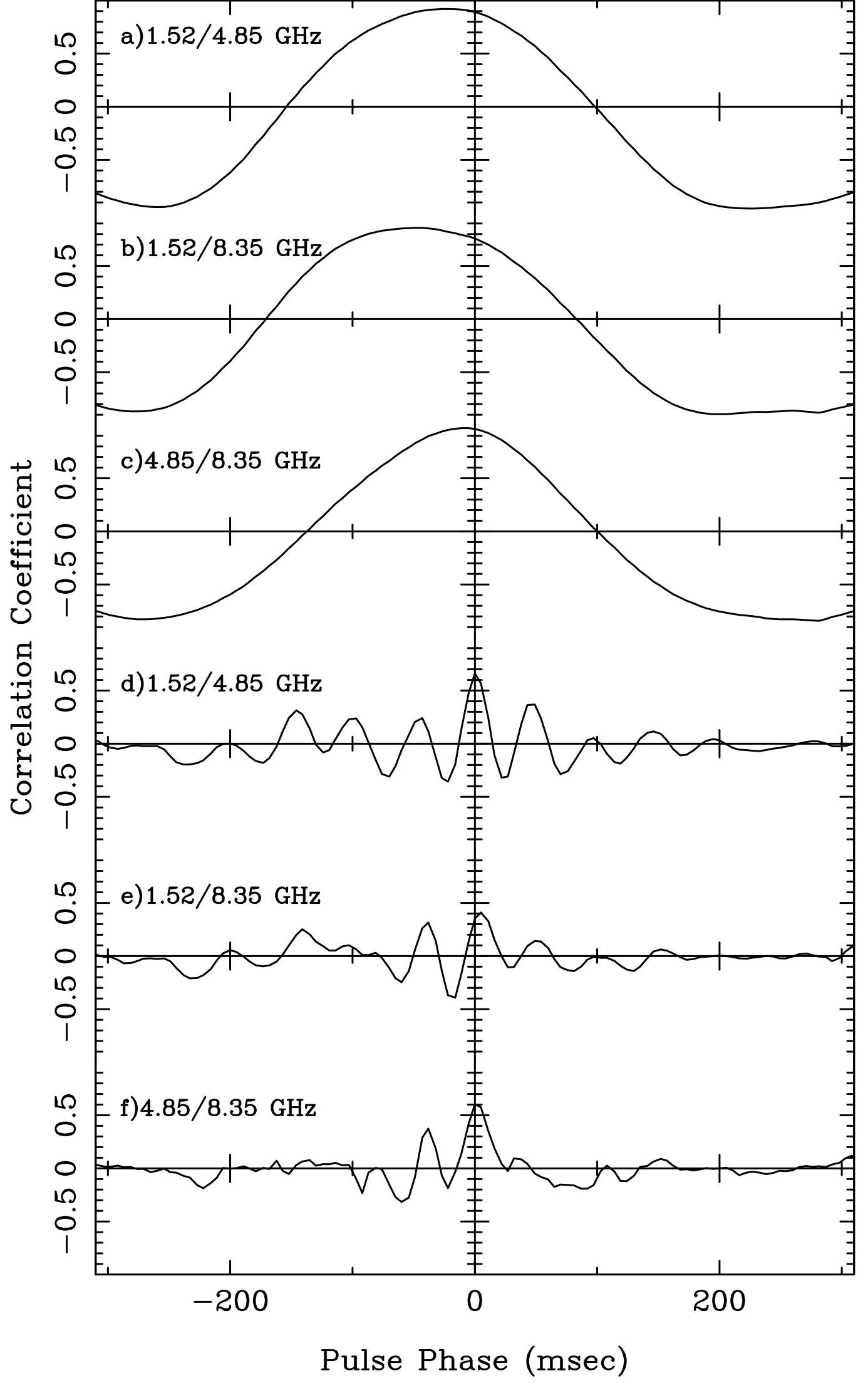}
    \caption{{\it Left:} 
    Detailed pulse profile structure of the main pulse component after averaging profiles obtained during the first ten days after the re-detection.  Panels a-c show respectively the normalised flux density (in arbitrary units) at 1.52 GHz recorded at Jodrell Bank and 4.85 and 8.35 GHz recorded at Effelsberg.  
    The profiles are aligned assuming a dispersion measure of 178.0 pc cm$^{-3}$.  Panels d-f show the same profiles after the application of a high-pass filter, which removed a Gaussian band of frequencies with half-power width of 13.8 Hz.  The flux density scale is the same as for panels a-c.  These reveal the presence of the periodic structure with period of about 50\,ms at all three radio frequencies. Peaks and troughs in intensity are seen to be approximately aligned at all three frequencies. 
    {\it Right:}
    Cross-correlation functions between the profiles presented in the left panels. Panels a-c are the cross-correlation functions between pairs of unfiltered profiles, showing little evidence of any high-frequency structure. Panels d-f present the cross-correlation functions between pairs of high-pass-filtered profiles. There is high correlation at zero delay between all 3 pairs.}
    \label{fig:corrs}
\end{figure*}

The periodicity was present in both JBO and Effelsberg data at all observing frequencies, as can be seen in Fig. \ref{fig:corrs}. Since the 1.52-GHz data from JBO shows that the structure is stable and fades at around MJD\,58468, we have averaged all available profiles obtained up to this date to improve the signal-to-noise ratios in each frequency band.  The good alignment of the peaks and troughs of the filtered profiles and the strong correlation coefficients at zero delay between the profiles in all 3 bands can be seen clearly.
No similar periodicity has been detected in any of the other profile components. 

\begin{figure}
    \centering
    \includegraphics[width=\columnwidth]{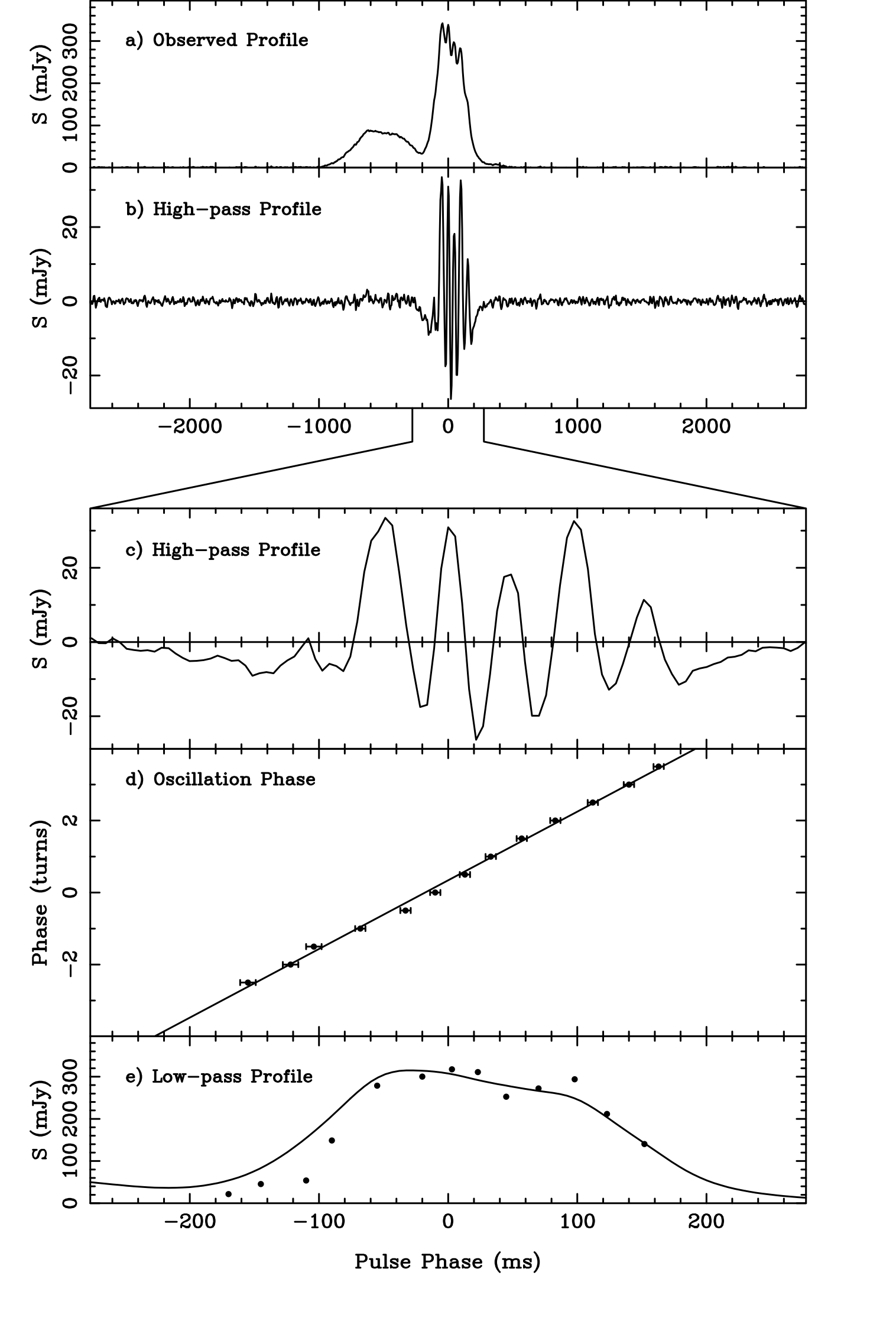}
    \caption{The 50-ms profile periodicity. 
    {\it a)} The integrated profile at 1.52\,GHz obtained by summing the first 10 observations. The flux density (S) is displayed across the whole pulse period.
    {\it b)} The same data as in a) after application of a high-pass filter as described in section 3.1, having a cut-off frequency of 9.2\,Hz.  Note that the high-frequency structure is confined to the region of the main pulse component. 
    {\it c)} The region around the main pulse component shown in b) expanded by a factor of 10 in time, in which about 7 cycles of the 50-ms periodicity are evident. 
    {\it d)} The phase of the oscillations as a function of time. The pulse phases of zero crossings (the points where the curve crosses the mean flux density of adjacent peaks and troughs of the pattern) are measured and assigned phases of $N$ and $N$+0.5 respectively for positive and negative slopes, where $N$ is an integral cycle number. The line is a weighted least-squares straight-line fit to the data and has a slope of 19.3(3)\,Hz, corresponding to a period of 51.6(9)\,ms. 
    {\it e)} The underlying low-pass profile of the main pulse component, obtained by subtraction of the high-pass structure of c) from the observed profile. The filled circular symbols represent the peak-to-peak amplitudes of the oscillations, estimated from c) as the magnitude of the difference in flux density between a peak (or trough) and the mean of its preceding and following troughs (or peaks). These values were multiplied by a factor of 6.0 before plotting. Note that the flux density of the oscillations approximately tracks the underlying flux density of the main pulse component profile.}
    \label{fig:newplot}
\end{figure}

In order to study the phenomenon in more detail, we have summed the data at 1.52\,GHz obtained from the first 10 observations, so obtaining the high signal-to-noise ratio profile of the whole pulsar period which is presented in Fig. \ref{fig:newplot}a. Fig. \ref{fig:newplot}b shows the profile after the application of a high-pass filter as described above, showing that the strong 50-ms periodic structure is confined to the main pulse component.  A 128-point Fourier analysis of 700\,ms of the profile centred on this component reveals a single, essentially unresolved spectral feature at close to 20 Hz. A similar analysis of 700\,ms of data spanning the precursor component is devoid of any clear spectral features.  In particular, the amplitude of any spectral feature at 20\,Hz must be less than $\sim$4$\%$ of that in the main-pulse component.

To study the fidelity of the periodic structure further, in Fig. \ref{fig:newplot}c we present an expanded view of the main pulse shown in Fig. \ref{fig:newplot}b, in which 6 or 7 cycles of the periodicity are seen. We have measured the "zero-crossing" times of the pattern, taken as being the pulse phase where the curve crosses the mean flux density of each adjacent peak and trough. 
We have plotted the accumulated phase of the oscillation, which increases by 0.5 for each crossing, against these times in Fig. \ref{fig:newplot}d.  This shows an excellent straight line, indicating that the period is essentially unchanging, right across the profile component. A formal straight-line fit to the 13 data points is shown, indicating that the mean period is 51.6(9)ms.

Fig. \ref{fig:newplot}e presents the underlying low-pass profile of the main pulse component, obtained by subtracting the high-pass profile of Fig. \ref{fig:newplot}c from the observed profile. The amplitude of the oscillations in Fig. \ref{fig:newplot}c clearly changes across the component, and we have estimated the peak-to-peak flux density for each peak and each trough in the oscillation and plotted these also in Fig. \ref{fig:newplot}e, after scaling by a factor of 6.  Remarkably, we note that the amplitude of the oscillation approximately follows that of the underlying main-pulse component, amounting to $\sim$17$\%$ of the underlying flux density.

\section{Timing}
Using the $\sim$daily 1.5-GHz observations collected with the Lovell telescope since the radio re-activation, we have obtained a phase-connected timing solution.  
Times of arrival (ToAs) were created by cross-correlating a standard profile with each observed pulse profile using the {\sc psrchive}\footnote{http://psrchive.sourceforge.net/} \citep{hot04} software package, and timing solutions were obtained using {\sc tempo2}\footnote{https://www.atnf.csiro.au/research/pulsar/tempo2/} \citep{hob06}. 
To analyse how the timing precision is affected by the choice of standard profile, we used a range of shapes based on the different pulse profiles from different epochs and compared the resulting timing precision. We found that the best overall fit (i.e. lowest RMS), which also resulted in the most evenly sized errors for the arrival times of separate epochs, was achieved when using a standard profile that was based on an average profile from a range of observing epochs, and meant that a single standard profile could be used throughout the entire timing process. 
Here we have used a standard profile that was created by aligning observations from 8--24 Dec 2018 ($\sim$8.5\,hours of data). 
The resulting timing solution is presented in Table \ref{tab:freq}. The only values fitted for were spin frequency ($\nu$) and its derivatives. The position of the pulsar was held fixed at the value reported by \cite{hel07} using VLBA observations. Since the timing analysis was performed at a single observing frequency also the DM was held fixed (at 178 \,pc\,cm$^{-3}$). 

\begin{figure}
	\includegraphics[width=\columnwidth]{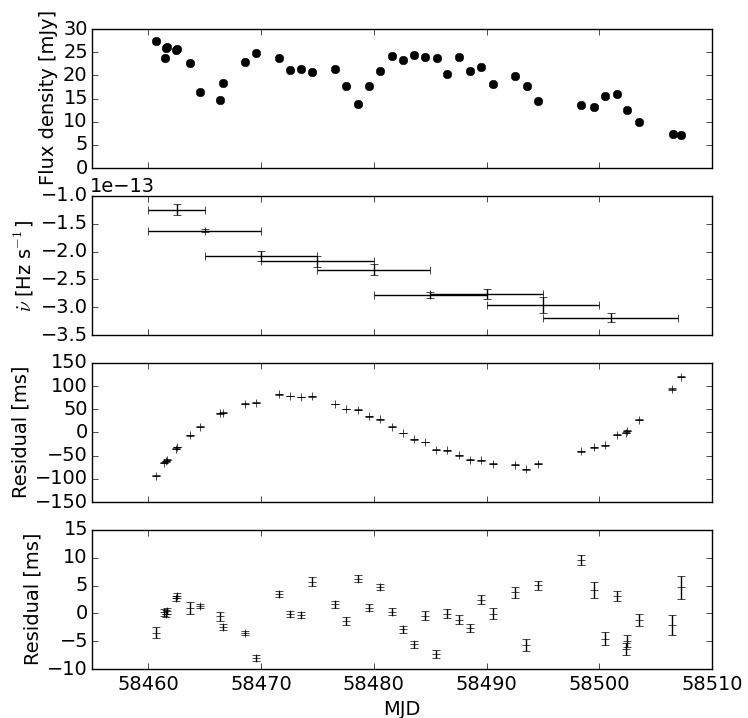}
    \caption{The top panel shows the flux density of XTE\,J1810--197 measured in data collected at the JBO from 8\,Dec\,2018 until 24\,Jan\,2019. 
    The second panel shows how the frequency derivative is changing with time, using $\sim$10 days of data in each fit, with a stride timescale of 5 days. In addition, we have included a first frequency derivative fit using only the first 5 days (8 observations), to show the great change in $\dot{\nu}$ in the beginning of the time span. 
    The third panel shows post-fit residuals in a fit to the entire JBO data set, using a timing model that includes $\nu$ and $\dot{\nu}$. In the bottom panel, $\ddot{\nu}$, $\dddot{\nu}$, and $\ddddot{\nu}$ have also been included. The parameters from the fits to the full data set are shown in Table \ref{tab:freq}.}
    \label{fig:residuals}
\end{figure}

The two lower panels in Fig. \ref{fig:residuals} show timing residuals for the entire 47-day data-set. The middle panel displays residuals from a fit to $\nu$ and $\dot{\nu}$, showing a clear cubic trend, with an RMS for the post-fit residuals of 0.01 pulse periods. 
Extending the fit to include $\ddot{\nu}$ gives an RMS of 0.0013 pulse periods, and adding also a $\dddot{\nu}$ to the fit gives an RMS of 0.0009 pulse periods. In both these cases, the resulting residuals are showing structure suggesting higher order frequency derivative terms are necessary to whiten the residuals. 
In the bottom panel frequency derivatives up to $\ddddot{\nu}$ have been included, resulting in an RMS of 0.0007 pulse periods. Adding additional frequency derivatives does not improve the fit significantly. In all cases, the RMS of the overall fit is larger than the errors on individual timing points, which is likely due to short term profile evolution.
The parameters from both timing fits in Figure \ref{fig:residuals} are shown in Table \ref{tab:freq}. 

The timing model suggests that $\dot{\nu}$ is changing rapidly, and we have performed a stride fit to the data to show the evolution of $\dot{\nu}$ with time. 
This is plotted in the top panel of Fig \ref{fig:residuals}, 
showing that $|\dot{\nu}|$ more than doubled within the first 15 days of observing, and that the rate of change decreased with time. The first measured value, $\dot{\nu}=-1.2(1)\times10^{-13}$\,Hz\,s$^{-1}$ at MJD=58460-58465, is similar to the values measured by \cite{cam16} just before the radio pulsations from the magnetar became undetectable in 2008 ($\sim$\,$-$9\,$\times10^{-14}$\,Hz\,s$^{-1}$ around MJD=54700). Over the next 47 days, $|\dot{\nu}|$ increased to a value similar to the first measured value for radio pulsations in 2006: $\dot{\nu} = -3.19(8) \times 10^{-13}$\,Hz\,s$^{-1}$ at MJD=58495-58507 compared to $\dot{\nu} = -3.3 \times 10^{-13}$\,Hz\,s$^{-1}$ around MJD=53830 \citep{camApJ663}. 
During the radio-quiet phase, XTE\,J1810--197 was still emitting X-ray pulsations and so it was possible to continue to time it \citep{pin16a,pin19}. The reported timing solutions suggest that the spin-down has been relatively stable during the quiescent phase, with $\dot{\nu}$\,=\,$-$9.2059(16)\,$\times10^{-14}$\,Hz\,s$^{-1}$ at MJD=55444 \citep{pin16err} and 
$\dot{\nu}$\,=\,$-$9.26(6)\,$\times10^{-14}$\,Hz\,s$^{-1}$ at MJD=58002.5 \citep{pin19}. 

\begin{table}
	\caption{Timing parameters of XTE\,J1810--197.} 
	\label{tab:freq}
	\begin{tabular*}{\columnwidth}{@{\extracolsep{\fill}} ll} 
		\hline
		\hline 
		Parameter & Value\\
		\hline
		\hline
		Right Ascension [hh:mm:ss] & 18:09:51.087\\
		Declination [dd:mm:ss]& -19:43:51.93\\
		DM [pc cm$^{-3}$] & 178.0\\
        Date range [MJD] & 58460--58507\\
        Epoch [MJD] & 58484\\
		$\nu$ [Hz] & 0.180458147(2)\\
        $\dot{\nu}$ [Hz s$^{-1}$] & -2.575(4)$\times$10$^{-13}$\\
        $\ddot{\nu}$ [Hz s$^{-2}$] & -5.5(4)$\times$10$^{-20}$\\
		$\dddot{\nu}$ [Hz s$^{-3}$] & 1.7(2)$\times$10$^{-26}$\\
		$\ddddot{\nu}$ [Hz s$^{-4}$] & 4.2(10)$\times$10$^{-32}$\\
		RMS of post-fit residual [period] & 0.0007\\
		\hline
		$\nu^*$ [Hz] & 0.180458125(1)\\
        $\dot{\nu}^*$ [Hz s$^{-1}$] & -2.49(2)$\times$10$^{-13}$\\
        RMS of post-fit residual$^*$ [period] & 0.01\\
        \hline\\
	\end{tabular*}
	{\bf Note:} The only values fitted were the spin frequency and its derivatives. Uncertainties are given in parentheses as 1-$\sigma$ errors on the last significant quoted digit.\\
	$^*$These values were obtained by fitting for $\nu$ and $\dot{\nu}$ only. The resulting residuals are shown in the middle panel of Fig. \ref{fig:residuals}. 
\end{table}

\begin{figure*}
    \includegraphics[width=17.5cm]{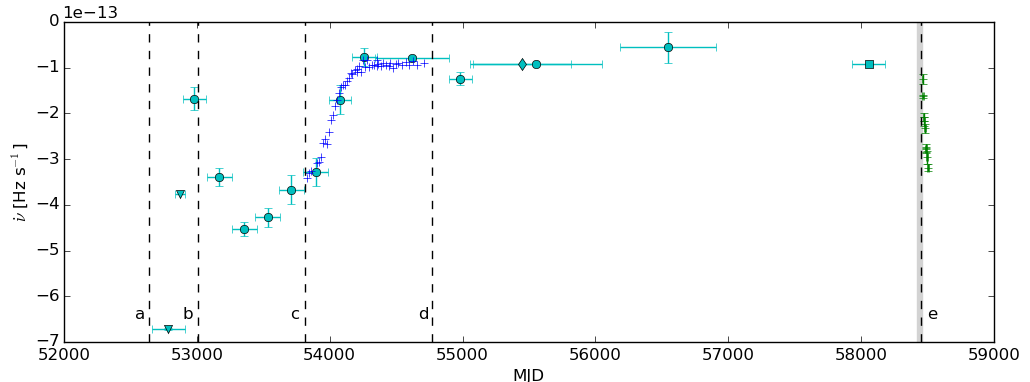}
       \caption{Spin-down history of XTE\,J1810--197, showing all published values of $\dot{\nu}$. The cyan-colored points are all from X-ray timing: triangles show values from Ibrahim et al. (2004), circles show values from Camilo et al. (2016), the diamond-shaped value is from Pintore et al. (2016b) and the square-shaped point is from Pintore et al. (2019). The blue crosses are estimated from the radio timing values in Camilo et al. (2016), and the green points (after MJD=58460) are from the radio timing presented here (shown in more detail in Fig. \ref{fig:residuals}). The black dashed line marked {\it a} represents the approximate time of the original X-ray outburst (Ibrahim et al. 2004), {\it b} shows the time of detection of a radio point source (Halpern et al. 2005), the area between {\it c} and {\it d} shows the previous active phase of radio pulsations (Camilo et al. 2016), the dashed line marked {\it e} shows the recent MAXI detection of increased X-ray emission (Gotthelf et al. 2019), and the grey region by {\it e} represents the time between our last non-detection in the radio band and our first observation in the radio-active state.}
   \label{fig:allF1}
\end{figure*}

\section{Discussion}
The on-set of radio pulsations in XTE\,J1810--197 was detected between 12--18 days after the X-ray emission was enhanced as observed by MAXI \citep{got19}. 
This is much closer in time than after the 2003 outburst, when the radio emission was only detected $\sim$a year after the X-ray enhancement \citep{hal05}. X-ray monitors missed the 2018 outburst, primarily due to the proximity of the magnetar's line of sight to the Sun during this time. This highlights the importance of radio monitoring of these sources even through radio-quiet phases, as without it the radio re-activation of XTE\,J1810--197 would have gone unnoticed. 

The flux density evolution of the magnetar, as measured in data collected at 1.52\,GHz with the JBO, is shown in the top panel of Fig. \ref{fig:residuals}. The variation seen is not uncommon for radio magnetars, which often show large and erratic flux density changes \citep[e.g.][]{lyn15,lev10}, and the magnitude of the variation is similar to what was observed for XTE\,J1810--197 in the first observations of radio pulsations after the 2003 outburst \citep{camApJ663}. 

The outburst and spin-down history of XTE\,J1810--197 is plotted in Fig. \ref{fig:allF1}. 
The observations of XTE\,J1810--197 collected during the first few hundred days after the 2003 outburst were unfortunately not sampled densely enough to resolve the details of the spin-down evolution directly after the burst. However, previously published values show highly variable mean $\dot{\nu}$-values for $\sim$500 days after the outburst (see Fig. \ref{fig:allF1}), followed by $\sim$1000 days of slowly decreasing $|\dot{\nu}|$, and finally reaching a minimum value $\sim$1500 days after the outburst. The spin-down remained relatively stable at this minimum value of $|\dot{\nu}|$ after the radio pulsations turned off in 2008. The last published $\dot{\nu}$-value ($\dot{\nu}=-9.26(6)\times 10^{-14}$\,Hz\,s$^{-1}$ at MJDs\,57932--58181) by \cite{pin19} is similar in magnitude to the first measured value in this work ($\dot{\nu}=-1.2(1)\times10^{-13}$\,Hz\,s$^{-1}$ at MJDs\,58460--58465). 
Over the next 47 days, $|\dot{\nu}|$ has increased by a factor of 2.6, with the most rapid increase occurring during the first 15 days. 

The rapid torque increase observed since the latest outburst is not unusual for radio magnetars. 
Similar $|\dot{\nu}|$ increases were seen both shortly after the 2017 outburst of PSR\,J1622--4950 \mbox{\citep[a factor of 7 in $\sim$60 days;][]{cam18}} and after the 2008 outburst of 1E 1547.0--5408 \citep[a factor of 4 in $\sim$30 days;][]{dib12}. 
Large torque variations were also observed after each of three outbursts of the Anomalous X-ray Pulsar (AXP) 1E 1048.1--5937 \citep{arc15}, however no similarly rapid torque increase was detected directly after the outbursts. 
Instead \cite{arc15} report on large $\dot{\nu}$ variations (of factors up to $\sim$7) starting $\sim$100 days after the outbursts and continuing for a few hundred days, before $\dot{\nu}$ stabilises again. 
Similar large $\dot{\nu}$ variations were seen after the 2007 outburst of PSR\,J1622--4950  \citep[a factor of $\sim$2 over a few hundred days;][]{lev12,sch17}, and after the 2003 outburst of XTE\,J1810--197 as described above. 
In both cases, the large variations were followed by a slow decrease in $\dot{\nu}$ until the pulsed radio emission turned off. 

A rapid torque increase was also observed after the 2016 X-ray outburst of  PSR\,J1119--6127 \citep{dai18}, which is a rotation powered pulsar displaying magnetar-like features. Directly after the X-ray outburst, the pulsed radio emission from PSR\,J1119--6127 turned off for $\sim$10 days \citep{bur16ATel-1,bur16ATel-2}. When it turned back on, $|\dot{\nu}|$ was observed to increase 
by a factor of $\sim$4 over
$\sim$30 days and then slowly return to the pre-outburst value after about 300 days \citep{dai18}. For PSR\,J1119--6127, the increase in spin-down torque seemed to be correlated with an increase in radio flux density, with the largest measured $|\dot{\nu}|$-value corresponding to the highest measured radio flux density value. As evident from fig \ref{fig:residuals}, we see no similar correlation for XTE\,J1810--197. 

In intermittent pulsars, $|\dot{\nu}|$ has been shown to be larger when the radio pulsations are on than when the pulsar is in its off state \citep{kra06B1931}. Similar trends are observed in both PSR\,J1622--4950 \citep{sch17, cam18} and XTE\,J1810--197, with the lowest value of $|\dot{\nu}|$ being observed just before the magnetars turned off in the radio band, and then increasing when radio pulsations are on.

Given the relative stability of $\dot{\nu}$ during the radio-quiet phase, it is possible to estimate what the spin period of XTE\,J1810--197 would be at the current observing epoch, if the spin-down parameters remained stable through the outburst. By extrapolating from the values of spin period and period derivative given in \cite{pin19} to the epoch of our first detection after the outburst, we found that the measured spin period (P\,=\,5.5414391(1)\,s on MJD 58461.5) is smaller than the one expected from the extrapolation (P\,=\,5.541463\,s) \citep{lyn18atel}. This could be interpreted as a glitch occurring in connection with the X-ray outburst. \cite{got19} calculates a possible glitch magnitude of $\Delta\nu/\nu=(4.52\pm0.15)\times10^{-6}$, which is similar to (although slightly larger than) the value of $4.32\times10^{-6}$, which follows from our timing measurements. Glitches of that size are not unusual for magnetars \citep{dib14}. 
On the other hand, if the spin down of the magnetar would have slowed down further between the last observation included in the solution in \cite{pin19} and our first observation after the outburst, the period change becomes consistent without the need to introduce a glitch. That would require a $\dot{\nu}\sim-6.03\times10^{-14}$\,Hz\,s$^{-1}$ between MJDs 58181--58461.5, which is not unreasonable given the spread of $\dot{\nu}$'s measured in the radio-quiet phase (see points between {\it d} and {\it e} in Fig. \ref{fig:allF1}).

The 50-ms periodicity seen in the pulse profile for about 10 days after it was redetected are extraordinary. Despite the usual variation of the magnetar pulse profiles (albeit less strong than seen a decade ago), it is remarkable that the structure persisted for more than a week, and was imprinted on the emission at all radio frequencies that we observed. We note in passing that this periodicity is unrelated to apparent similar features seen in the position angle swings observed at 8.45 GHz by \cite{kra07}. These were evidently caused by power-line signal of 50 Hz leaking into the data. The oscillations here have a characteristic frequency of 20 Hz and are seen at different frequencies and different telescopes at the same time. 
No such persistent periodicity has been reported in the pulse profile of any other radio pulsar. The constancy in phase of the periodic structure relative to the main pulse profile indicates that the periodicity is not a time modulation of the emitting source, but must be due to a periodic structure in the radiation beam pattern that sweeps across the Earth as the pulsar rotates. This demands an (magnetic inclination angle dependent) angular scale of 0.052/5.54$\times$360 = 3.4 degrees for the observed 17$\%$ regular modulation pattern which is superimposed on the 300-ms wide (20-degree) main component beam. In two dimensions, one can envisage this 20-degree component beam with modulating stripes across it having an angular scale which would depend upon the angle between the passage of the line of sight to the Earth and the direction of the stripes. We note that there is no such modulation of the flux density in the precursor component or anywhere else across the pulse period.
In normal radio pulsars which have multiple components, it is usual to invoke a beam which consists of multiple concentric annular cones, centred on a magnetic pole. The observed components occur as the cones pass over the line of sight to the Earth. Such an explanation is
difficult to invoke in the case of the periodicity in the beam of XTE\,J1810--197, because of the observed constancy of the component separation across the beam revealed in figure 5d.
We believe that the stripes in the beam pattern must arise in some way from a stable structure on the surface of the neutron star at the base of the magnetic field lines hosting the emitting particles for the radio component. This could for instance arise from a frozen-in wave pattern of surface ripples of height, temperature or magnetic field. This pattern might somehow modulate the emission of particles which feed the emission engine. The mechanism of such emission variation is probably the same as the one that gives pulsars their individual complex pulse profiles. 
Such a pattern is reminiscent of surface waves in the neutron star crust \citep{pir04}, which may be produced as high-spherical-degree non-radial oscillations \citep{cle04}, but we note that those theories describe oscillatory effects rather than a static structure, which we believe is necessary.

Quasi-periodic oscillations of X-ray emission during magnetar flares are not uncommon \citep[e.g.][]{str05,isr05}, with frequencies of similar magnitude to the one observed here.  
However, the time scales of the phenomena are different, with X-ray quasi-periodic oscillations lasting only a few seconds and the radio oscillations seen here lasting at least 10 days.

The pulse profile observed from the XTE\,J1810--197 in May to August 2006 showed dramatic variations in fluxes, shape and positions of different components across all the frequency bands on timescales as short as a few days \citep{kra07,camApJ663}. Comparison of pulse profiles on longer timescales also show variations, but with the lower cadence the nature of the timescale of the variability is less easy to judge \citep{cam16}. The pulse variations seen so far from the source have been significantly less dramatic, on timescales from hours to months, than seen in 2006. It will be interesting to see whether this changes as we get further from the event which triggered the radio emission again, i.e. at a time after the burst similar to the 2006 observations. Perhaps it also might show some correlation with a change from a negative to positive rate of change of the frequency derivative, i.e. as it returns to the quiescent value.

\section*{Acknowledgements}
 Pulsar research at Jodrell Bank Centre for Astrophysics and access to the Lovell telescope is supported by a Consolidated Grant from the UK's Science and Technology Facilities Council. BWS acknowledges funding from the European Research Council (ERC) under the European Union's Horizon 2020 research and innovation programme (grant agreement No. 694745). 
 GD, RPE, RK and MK acknowledge financial support by the European Research Council for the ERC Synergy Grant BlackHoleCam under contract no. 610058. This work was based on observations with the 100-m telescope of the Max-Planck-Institut f{\"u}r Radioastronomie at Effelsberg.



\bibliographystyle{mnras}
\bibliography{1809}

\bsp	
\label{lastpage}
\end{document}